\documentstyle [12pt]{article}
\newcommand{\be}{\begin{equation}}
\newcommand{\ee}{\end{equation}}
\begin{document}
\begin{center}
{\bf Quantum Transmission in Disordered Insulators: \\
Random Matrix Theory and Transverse Localization}

\vspace{5ex}  Yshai AVISHAI$^{(a,b)}$ and Jean-Louis PICHARD$^{(a)}$\\
\vspace{2ex} (a)  C.E.A., Service de Physique de l'\'Etat Condens\'e. \\
 Centre d'\'Etudes de Saclay,  91191 Gif sur Yvette Cedex,    France.\\
\vspace{2ex} (b) Department of physics, Ben Gourion university, Beer Sheva,
Israel.\\
\vspace{3ex} and\\
\vspace{2ex}  Khandker A. MUTTALIB\\
\vspace{3ex} Physics Department, University of Florida, Gainesville, Fl 32611,
U.S.A.

\end{center}

\vspace{3ex}

\noindent{\bf Abstract:} \\

{\it We consider quantum interferences of classically allowed
or forbidden electronic trajectories in disordered dielectrics.
Without assuming a directed path approximation, we represent a
strongly disordered elastic scatterer by its transmission matrix
${\bf t}$. We recall how the
eigenvalue distribution of ${\bf t.t}^{\dagger}$ can be obtained
from a certain ansatz leading to a Coulomb gas analogy at a
temperature $\beta^{-1}$ which depends on the system symmetries.
We recall the consequences of this random matrix theory for
quasi--$1d$ insulators
and we extend our study to microscopic three dimensional models
in the presence of transverse localization. For cubes of size
$L$, we find two regimes for the
spectra of ${\bf t.t}^{\dagger}$ as a function of the localization
length $\xi$. For $L / \xi \approx 1 - 5$, the eigenvalue spacing
distribution
remains close to the Wigner surmise (eigenvalue repulsion). The
usual orthogonal--unitary cross--over is observed for {\it large}
magnetic field change $\Delta B \approx \Phi_0 /\xi^2$ where
$\Phi_0$ denotes the flux quantum. This field reduces the
conductance fluctuations and the average log--conductance (increase
of $\xi$) and induces on a given sample large magneto--conductance
fluctuations of typical magnitude similar to the sample to sample
fluctuations (ergodic behaviour). When $\xi$ is of the order of the
lattice spacing ($L/\xi \gg 5-6$), the eigenvalue repulsion is weaker
and the removal of a time reversal symmetry has a more negligible
role. In those two regimes, the sample to sample fluctuations of
the log--conductance are close to a normal one--parameter
distribution, and $\xi^{-1}$ depends linearly on the disorder parameter.
In these microscopic models, the Fermi energy dependence of the
conductance is very similar to the one recently observed in small
GaAs:Si wires.}

\vspace{10ex}

Submitted for publication in Journal de Physique.
\newpage

\vspace{3ex}

\noindent{\bf R\'esum\'e:} \\

{\it Nous consid\'erons les interf\'erences quantiques de
trajectoires electroniques classiquement permises ou interdites
dans des di\'electriques d\'esordonn\'es. Sans faire une approximation
de chemins dirig\'es, nous repr\'esentons un diffuseur \'elastique
tr\`es d\'esordonn\'e par sa matrice de transmission ${\bf t}$.
Nous rappelons comment la
distribution des valeurs propres de ${\bf t.t}^{\dagger}$ peut
\^etre obtenue \`a partir d'un certain ansatz conduisant \`a une
analogie avec un gaz de Coulomb \`a une temperature $\beta^{-1}$
qui depend des sym\`etries du syst\`eme. Nous rappelons les
cons\'equences de cette theorie de matrices aleatoires pour des
isolants quasi--$1d$ et nous \'etendons notre \'etude \`a des mod\`eles
microscopiques tridimensionnels o\`u la localisation transverse est
pr\'esente. Pour des cubes de taille $L$, suivant la valeur
de la longueur de localisation $\xi$, nous trouvons deux regimes
pour les spectres de ${\bf t.t}^{\dagger}$.
Quand $L/\xi \approx 1-5$, la distibution
des espacements entre valeurs propres reste proche de celle de
Wigner (repulsion des valeurs
propres). Le cross--over habituel entre les cas orthogonal et unitaire
a lieu pour un champ magnetique appliqu\'e \'elev\'e $\Delta B \approx
\Phi_o/\xi^2$ o\`u $\Phi_o$ est le quantum de flux. Ce m\^eme champ
r\'eduit les fluctuations de conductance $g$ ainsi que la moyenne de
$log(g)$ (augmentation de $\xi$) et induit sur un echantillon donn\'e
de grandes fluctuations de la magnetoconductance d'une ampleur comparable
\`a celles d'\'echantillon \`a \'echantillon (comportement ergodique).
Quand $\xi$ est de l'ordre de la maille du r\'eseau, ($L/\xi >>5-6$),
la r\'epulsion des valeurs propres est plus faible et l'effet du
champ magnetique cesse d'\^etre important. Dans ces deux r\'egimes,
les fluctuations d'echantillon \`a echantillon de $log(g)$ sont
bien d\'ecrites par une gaussienne \`a un param\`etre et $\xi^{-1}$
varie lin\'eairement en fonction du param\`etre de d\'esordre. Dans ces
ces mod\`eles microscopiques, la dependance en fonction de l'energie
de Fermi de la conductance est tr\`es comparable \`a celle recemment
observ\'ee sur de petits fils de GaAs:Si. }

\newpage

\vspace{3ex}
\noindent{\bf 1. Introduction}
\vspace{1ex}

  When one has to tackle the problem of electronic quantum transport,
one of the usual microscopic starting points is the familiar Anderson
tight--binding Hamiltonian:
\be
H=\sum_{i} \epsilon_i a_i^{\dagger} a_i +\sum_{<i,j>} h_{ij} a_i^{\dagger}a_j
\ee
where the $\epsilon_i$ are the random site energies and the $h_{ij}$
represent the nearest
neighbor couplings or transfer terms. For instance, the $\epsilon_i$
can be random with a rectangular distribution of width $W$ (Anderson model)
or can take at random two values $\epsilon_a$ or $\epsilon_b$
(binary alloy model). The $h_{ij}$ can take for simplicity a uniform value $h$
if $i,j$ are nearest neighbors, 0 otherwise.
 Perturbations could be done when
one of the two terms of $H$ (random substrate electrostatic potential
or kinetic energy) is negligible compared to the other. The scaling
theory of localization~[1] assumes the existence of a universal
$\beta(g)$--function of conductance $g$ relating those two limits
through the mobility edge in three dimensions.

 For disordered metals whose quantum conductance $g$, measured in unit
of $e^2/h$, is larger than one, the conventional perturbation
approach starts from plane--wave like states which are only weakly
perturbed by the scattering caused by the random substrate ($W/h \ll 1$).
In this limit, electronic transport essentially results from
{\it quantum interferences of classically allowed diffusive
paths} and quantum theory is close to its semi--classical limit.

(i) We know  for the three different possible symmetries
(orthogonal, unitary and symplectic cases):
the four loop order expansions~[2] in $2+\epsilon$ dimensions of the
scaling functions $\beta(g)$. One gets after integration the
weak localization corrections to Boltzmann conductance, i.e. how $<g>$
depends on the sample size $L$, the brackets denoting the ensemble average
value.

(ii) The sample to sample conductance fluctuations around the ensemble
average essentially  satisfy gaussian distributions whose variance
depends only on the basic symmetries and slightly
 on the dimensionality of the
system~[3] (Universal Conductance Fluctuations).

(iii) The conductance fluctuations induced on a given sample
by the variation
of an applied magnetic field or of the Fermi enegy are of similar
magnitude~[4] (ergodic behaviour).  An applied magnetic
field  removes time reversal symmetry,
suppressing the weak (anti) localization correction for $<g>$ in the
absence (in the presence) of spin--orbit scattering, halving
the variance of the universal conductance fluctuations.
The cross-over field is in general small in metals, corresponding
to the application of a flux quantum $\Phi_0$ through the temperature
dependent area where electrons keep their quantum coherence.

(iv) The conductance fluctuation yielded by a local reorganization of the
substrate potential (single impurity move) have been understood mostly
from known results on random walks, yielding a possible explanation for
universal amplitude of observed $1/f$--noises~[5].

 For Anderson insulators, $W/h \gg 1$ and it is better to perturb the
point--like--localized eigenstates of the random term in (1) by the kinetic
transfer term. The perturbation removes the possible degeneracies of the
$\epsilon_i$ (which is huge in the binary alloy model) and the eigenstates
of (1) are still localized over domains of typical size $\xi$ which
can extend over many impurities in the vicinity of the mobility edge.
The Green function $G(E)$ between two sites i and j can be expanded in
powers of the hopping integral $h$
\be
<i|G(E)|j> =\sum_{\Gamma} \prod_{i_{\Gamma}}{ h e^{iA} \over
E-\epsilon_{i\Gamma}}.
\ee
Here we have also introduced a magnetic vector potential $A$ for generality.
The paths $\Gamma$ going from i to j are in this case a priori highly non
classical, and electronic transport mainly results from
{\it classically forbiden trajectories which
interfere due to the random sign of the denominator and to the vector
potential}. When {\it all} the
factors $h/(E-\epsilon_i) \ll 1$, the shortest directed paths $\Gamma_d$
need only to be considered~[6] and electron trajectories with self
intersections
have no significant contribution in this limit:
\be
<i|G(E)|j> \approx h^L \sum_{\Gamma_d} \prod_{i_{\Gamma_d}}
{ e^{iA} \over E-\epsilon_i}
\ee
The combinatorics of the shortest directed paths $\Gamma_d$ of length $L$
can be simply obtained if some cases (e.g if the sites
$i$ and $j$ stand at the opposite corners of a square lattice).
More generally, these directed paths relating two points separated by
$L$ have transverse fluctuations extending over $(L \xi)^{\zeta}$.
Assuming standard diffusion, one gets $\zeta=1/2$ but a more detailed
analysis shows that dominant paths are ``superdiffusive'', yielding
different values~[7] for $\zeta$. One finds that:

(i) The localization length $\xi$ characterizing the decay of the propagator:
\be
<log | <i| G(E) |j> |^2 > \approx L(\rho - \xi_0^{-1})
\ee
depends on the disorder parameter $W$ through a local contribution
$\xi_0=[log(W/h)]^{-1}$ which comes from the exponentially small hopping
term $[h/W]^{L}$ in (3) (binary alloy model with $\epsilon_i= ^{+}_{-} W$
at $E=0$). $\rho_0$ is a global contribution embodying the $W$--independent
quantum interferences of the directed paths in this particular model~[8].
$\xi$ increases as a function
of an applied field $B$ as $B^{1/2}$ with a universal factor, giving rise
to a positive magnetoconductance~[9]. This sign of the magnetoresistance
is not reversed by the presence of a strong spin--orbit scattering~[10],
as it is in disordered metals.

 (ii) The sample to sample conductance fluctuations are lognormally
distributed, but with anomalous size dependence of the fluctuations.
$var(log(g)) = C. L^{2\omega}$, where the prefactor $C$ does not depend on
the mean $< log(g)> \propto L$, $\omega = 1/3$ in two dimensions
and $\omega \approx 1/5$ in three dimensions~[7-8].
The distribution of $log(g)$ then contains more than a single parameter.

 (iii) The application of a flux quantum $\Phi_o$ through the ``cigare--shape
domain'', which contains the forward scatterered trajectories, yields a
fluctuation $\delta(g) \approx g$ much smaller than
the typical
sample to sample fluctuations~[11] (non ergodicity). No other characteristic
magnetic
field scale appears if self--intersecting trajectories are neglected.

 (iv) The effect of the motion of a single scatterer has not been addressed in
directed path theories. Broad distributions of the log--conductance
with a quasi--universal variance have been numerically observed
without a directed path approximation~[12].

 The restriction of quantum interferences to this class of directed
trajectories is usually made in ad hoc models (binary alloy distribution
at an energy far from the $\epsilon_i$) where one assumes
that the path contribution is always an exponentially decreasing
function of its length.  This does not take into account situations
where small denominators could compensate the exponentially small
value of $h^L$. Such quantum resonances necessarily occur in a device
where carrier concentration can be varied by capacitive technics.
This approximation does not describe also more general cases where
transport involves complicated interferences between classically
forbidden and allowed trajectories (vicinity of the mobility edge).
In an Anderson model with a broad distribution of $h/\epsilon_i$,
there is a finite probability to have clusters where the electron can
diffuse easily, be backscattered and selfintersects its own trajectory
before having to tunnel through large energy barriers. Such
metallic clusters can yield non trivial magnetic field or spin--orbit
scattering dependence in the vicinity of the mobility edge.

  An alternative non perturbative theory for the analysis of quantum
transmission in insulators is provided by a maximum entropy ansatz~[13]
using the transfer matrix $M$. This approach
quantitatively describes quantum transmission through quasi--$1d$
wires over length scale smaller or {\it larger} than their
localization lengths, leading us to predict novel
universal symmetry breaking effects~[14]. We are interested to know its
degree of validity outside quasi--one dimension in the presence of a
strong transverse localization, and to compare its predictions from
perturbation theory restricted to directed paths.
 We do not restrict ourself to the limit of an
elementary localization of the conduction electrons around
randomly located donors or acceptors, i.e. to a purely geometrical
disorder, but we consider also systems closer to a mobility edge where
localization results from {\it complex interferences between classically
allowed or forbiden trajectories}. In this later case, it is likely
that the relevant quantum interferences do not only involve simple
forward scatterered directed paths, and that the effect of a strong
spin--orbit coupling or of a large applied magnetic field have to be
revisited~[15, 16].  Being outside a simple perturbative limit, we rely on
numerical simulations which include all
possible path contributions without any a priori simplification.
We use as microscopic models either a random quantum wire network
or an Anderson
model with rectangular distribution of $\epsilon_i$.

 This work belongs to a series of three papers devoted to quantum
interferences in disordered insulators. One of them is theoretical~[17]
and analyzes how the purely quantum interferences which we consider are
truncated by activated process at low temperature. The second~[18] is
experimental and gives the energy and magnetic field dependence of
the conductance. One could then compare the results of our simulations
to their results and note a striking (at least qualitative) similarity.
The difficulty for a more quantitative comparison is due to the activated
processes.

We recall in sections 2 and 3 the maximum entropy description of
quantum transmission and its implications for quasi--one dimensional
insulators (typical conductance and sample to sample fluctuations,
symmetry breaking effects). We study in section 4 the modification
of the physical constraint used in this maximum entropy description
for three dimensional insulators. The persistence of the short range
repulsion in presence of transverse localization is shown in section 5
through a study of the eigenvalue nearest spacing distribution of
${\bf t.t}^{\dagger}$, though the long range interactions are weaker
than usual random matrix behaviour (study of the spectral rigidity).
In section 6, the sample to sample fluctuations of $\ln(g)$ are studied
in three dimensions for large disorder. We show that $\xi^{-1}$ varies
as a function of the disorder parameter linearly from the mobility edge
to the strongly localized regime.
The effects of a variation of the applied magnetic field and of the
Fermi energy are considered in section 7 and 8. We compare our results
for the eigenvalues of ${\bf t.t}^{\dagger}$ with those which are
understood for the system Hamiltonian (section 9), and with the
experimental results of Ref. 18 (section 10).

\vspace{3ex}
\noindent{\bf 2.  Maximum entropy description of quantum scatterers}
\vspace{1ex}

  We consider a $N$--channel quantum scatterer that we represent
by its transfer matrix $M$. This matrix gives the electronic flux
amplitudes at the right hand side of the scatterer in terms of the
flux amplitudes standing at its left hand side. The statistical
ensemble for a matrix $X$, related to $M$ by the relation
\be
X=(M^{\dagger}.M+(M.M^{\dagger})^{-1}-{\bf 1})/4,
\ee
is chosen~[13] as the most random matrix ensemble given the density
$\sigma(\lambda)$ of the eigenvalues of $X$.
This statistical ensemble of maximum information entropy for $X$
implies that the $X$--eigenvectors are statistically
independent of the eigenvalues and are distributed totally at random.
We denote by ${\bf t}$ the transmission matrix at the Fermi energy.
Using a two--probe Landauer formula we express the
conductance $g=2. tr ({\bf t.t}^{\dagger})$ as a linear statistics of
the eigenvalues $T_a$ of ${\bf t.t}^{\dagger}$ or the eigenvalues
$\{\lambda_a\}$ of $X$
via the relations:
\be
g=2.\sum_{a=1}^N T_a=2.\sum_{a=1}^N {1 \over {1+\lambda_a}},
\ee
the factor two coming from a possible spin or Kramer degeneracy.
Sometimes, we refer to the eigenvalues  $\{\lambda_a\}$ of $X$
as the radial parameters of $M$,
since they can be also introduced through a natural ``polar'' decomposition
of the transfer matrix.

 The joint probability distribution $P(\{\lambda_a\})$
can be easily calculated from this maximum entropy ensemble.
One gets an expression formally identical to the Boltzmann weight of a
set of classical charges free to move on a half line (the positive
part of the real axis) at a well--defined
``temperature''equal to $\beta^{-1}$:
\be
P(\{\lambda\}) \propto \exp-\beta.H(\{\lambda_a\}),
\ee
and governed by an hamiltonian:
\be
H(\{\lambda_a\})=-\sum_{a<b}^N \ln |\lambda_a-\lambda_b|+\sum_{a=1}^N
V(\lambda_ a).
\ee
The confining potential is given in terms of the input density
$\sigma(\lambda)$
    by
\be
V(\lambda)=\int_{0}^{\infty} \sigma(\lambda') \ln |\lambda - \lambda'|
d\lambda'.
\ee

 In usual random matrix theories, this is the classical Coulomb gas analogy
which introduces a temperature, a one--body potential and a two--body
interaction.

(i) The temperature is given by the fundamental symmetries which leave the
system
invariant (time reversal symmetry, spin rotation symmetry), $\beta=1,2,4$
respectively for the orthogonal, unitary and symplectic case.
In the absence of spin--orbit scattering, $\beta=1$ without applied magnetic
field
and becomes equal to 2 if the applied magnetic field is large enough (removal
of
time reversal symmetry).

(ii) The confining potential $V(\lambda)$ can be adjusted in order to attract
most
of the $\lambda_a$ near the origin, yielding a large value for $g$ (metal), or
on
the contrary $V(\lambda)$ can be very weakly attracting, leaving the
``charges''
spread out along the positive real axis, yielding $g \ll 1$ (insulator).
In other words, the spectrum of $X$ depends on the different physical
parameters (Fermi momentum $k_f$, disorder measured
by $l$, system size $L$...) through a simple one--body
potential $V(\lambda)$ in this maximum entropy ansatz.

(iii) As the value of $\beta$, the crucial logarithmic interaction between
the $\lambda_a$ is obtained from the invariant measure of the space
of matrices $X$ which are invariant under certain symmetries that each member
of the statistical ensemble must satisfy.

 The consequences of the ansatz $(8)$ on the spacing distribution of successive
radial parameters and on their two point correlation functions have been
carefully investigated~[19-20] for disordered conductors where the
$\lambda_a$ are
in part accumulated close to the origin. Detailed studies confirm both
the presence of the infinite range logaritmic interactions and the change
of temperature
associated with symmetry breaking effects {\it when the sample shape is nearly
quasi--one dimensional}. For higher dimensions ($d=2$  and $3$), the
transverse
diffusion slightly reduces the rigidity of the $\lambda$--spectra for high
transmission or reflection (edges of the spectra), albeit the random matrix
ansatz remains a good approximation.

  In the quasi--$1d$ limit, the ansatz (3) is consistent with a diffusion
equation~[13] which implies the quantitatively correct behaviours for the
average and the variance
of $g$ in the metallic limit, if the input density $\sigma(\lambda)$
satisfies a certain integro--differential equation. More important for the
questions investigated here, this ansatz remains valid when the sample length
is larger than the quasi--$1d$ localization length, establishing the
validity of
this ``Coulomb gas model'' in the presence of longitudinal localization. In
other
words, the Coulomb gas model remains essentially stable with respect to matrix
multiplication for a fixed number $N$ of channels.

\vspace{3ex}
\noindent{\bf 3. Longitudinal localization without transverse localization
(quasi--$1d$ limit)}
\vspace{1ex}

 We summarize in this section previous results obtained from the global
ansatz~[13-21]
and based on an analytic expression for $\sigma(\lambda)$ yielded from
the diffusion equation~[22] valid in the quasi--$1d$ limit.
The analysis is easier in terms of the variables $\{\nu_a\}$
related to the $\{\lambda_a\}$ through
\be
\lambda_a={(\cosh(\nu_a) -1) \over 2}.
\ee
In the large $N$--limit, one gets for the $\nu$--density
\be
\sigma(\nu)={2l \over NL} \ \ \ 0 \leq \nu \leq 2L/l
\ee
and zero elsewhere which gives for the confining potential:
\be
U(\nu)={Nl \over 2L} \int_0^{2L/l} \ln | \cosh(\nu) -\cosh(\nu ')|d\nu '
\approx
    {N.l \over 4L} \nu^2.
\ee
We write again the distribution $P(\{\nu_a\})$ as a Boltzmann weight at a
temperature $\beta^{-1}$ of a classical system, getting for the hamiltonian
\be
H(\{\nu_a\})=-\sum_{a<b}^N V_2(\nu_a,\nu_b)+\sum_{c=1}^N V_1(\nu_c).
\ee
 The two body interaction is now:
\be
V_2(\nu_a,\nu_b)=\ln |\cosh(\nu_a)-\cosh(\nu_b)|,
\ee
and the one body potential contains a $\beta$--dependent term yielded by
the change of variable:
\be
V_1(\nu_c)=-{1 \over \beta} \ln (\sinh(\nu_c))+ U(\nu_c).
\ee
 However, the quasi--$1d$ localized limit is characterised by a particular
behaviour
of the spectral density: increasing the system size $L$ for a fixed number $N$
of
channels, we end up to have:
\be
1 \ll \nu_1 \ll \nu_2 \ll \ldots \ll \nu_N
\ee
which allowes us to simplify $V_2(\nu_a,\nu_b)$, since
$\ln|\cosh(\nu_a)-\cosh(\nu_b)| \approx \max(\nu_a,\nu_b)$, to get
approximately
independent gaussian fluctuations of $\nu_a$ centered at
\be
\nu_a^0=(a-1+{1\over\beta}){2L/Nl}
\ee
and of variance $2L/(\beta Nl)$.
 Since $\nu_1=2L/\xi$, we see that the localization length is proportional to
$\beta$:
\be
\xi=\beta Nl.
\ee
The $N$--dependence of $\xi$
is characteristic of the quasi--$1d$ limit where tranverse localization
is absent.  Change of $\xi$ and change of the magnitude of the
conductance fluctuations by a symmetry breaking effect are the two sides of
the same phenomenon, since $\xi$ is defined from the ensemble average
$<log(g)> \approx 2L/\xi $ and not from $<g>$. The
characteristic cross--over fields reducing the magnitude of the conductance
fluctuations and increasing $\xi$ in the absence of spin--orbit scattering
are typically larger in insulators than in conductors. The
change from orthogonal to unitary symmetry occurs
when a flux quantum $\Phi_0$ is applied through the phase coherent domain
in a conductor (temperature dependent) and through the localization
domain for an insulator (temperature independent). The corresponding
fields $B^*$ are of order of $\approx 10 - 100$ gauss in a conductor
at very low temperature (a few mK) and  $\approx 0.5 - 2 $ teslas for a
large localization length insulator. For insulators with
strong spin--orbit scattering , if one assumes that the removal of the
Kramer degeneracy for $\nu_1$ by an applied magnetic field $B$ introduces
a splitting which is negligible compared to the large value of $2L/\xi$,
the transition $\beta=4 \rightarrow \beta=2$ essentially halves the
value of $\xi$. The existence of a positive magnetoresistance for insulators
with strong spin--orbit scattering is in this theory the counterpart
of the suppression of weak--anti--localization in conductors, in agreement
 with some experiments~[16] and in disagreement with directed path
theories~[10].

 The second important result that we have derived concerns the sample to
sample fluctuations of the logarithm of the conductance since
\be
-\log (g) \approx 2L/\xi = \nu_1
\ee
in this quasi--$1d$ insulating limit. We obtain a normal distribution of
$\log(g)$ with
\be
var(log(g)) = -< log(g)>
\ee
 Note that contrary to approximations based on directed paths, $log(g)$
has a one--parameter distribution.

 In summary, the above results indicate that the validity of a ``Coulomb gas
model''
for the $\{\lambda_a\}$ in the quasi--$1d$ limit is not at all equivalent
to the validity of Wigner--Dyson statistics since the spacings between
successive $\lambda_a$
become so exponentially large that the two--body interaction is reduced to an
effective one body potential. In particular, this density effect is
responsible for the non-universality of $var(g)$ in insulators.
But one cannot check easily in this limit the
underlying presence of the logarithmic interaction.

 If we consider now strongly disordered $L*L*L$ cubes, the increase
of their size
$L$ will increase the number $N=(k_f.L)^{d-1}$ of channels in addition
to the longitudinal length. Real $3d$--insulators are then more likely
than quasi--$1d$ systems to be characterised by an exponentially large
value of $\lambda_1$ in the large $L$--limit without having exponentially
large intervals between successive $\{\lambda_a\}$. Does the Coulomb
gas analogy  which is at least an excellent approximation
in the absence of tranverse
localization, remains of some interest in a strong disorder limit?
Do two successive $\lambda_a$ still repel each other
when their spacing is not exponentially large, as implied
by the ansatz? Does the coulomb gas ``temperature'' take the classical
values implied by symmetry considerations? This is the first issue that
we have to check in our numerical study.

\vspace{3ex}
\noindent{\bf 4. Density and confining potential in the presence
of transverse localization ($3d$--limit)}
\vspace{1ex}

  We have calculated series of $\{\lambda_a\}$ for different realizations
of a microscopic model. We have used a $3d$--network of disordered
quantum wires as explained in appendix 1. Each link $(i,j)$ of this network
is characterized by a one--dimensional uniform electrostatic potential
$\epsilon_{i,j}$, which is taken at random with a rectangular distribution
of width $W$.  Real (propagating mode) or imaginary
(decaying mode)  wave--vector $k_{i,j}=(E-\epsilon_{i,j})^{1/2}$ characterizes
the electronic wavefunction along the link $(i,j)$. The matching conditions
at each intersection satisfy the basic symmetry and conservation laws of
quantum mechanics.
We have alternatively used the Anderson model (1) with the same distribution
for the random potential $\epsilon_i$, obtaining similar conclusions.
(See also another recent numerical study of $3d$ Anderson models by Markos
and Kramer~[23]).

In figure 1, the density $\sigma(\nu)$ is given for $3d$--cubes close to
the mobility edge (W=4) and for larger disorder (W=7). Contrary to
disordered conductors
where this density is mainly concentrated near the origin~[20], one can
see that for large values of $W$, $\sigma(\nu)$ is negligible in the
vicinity of the origin, indicating exponentially small
transmission. In figure 2, we show the ensemble averaged positions $<\nu_a>$
which are not widely separated for a dielectric cube of size $L$,
contrary to the quasi--$1d$ limit. One can note however that the spacings
are larger near the lower edge of the spectra than in the bulk.
We note that $g<1$ both for $W=4, \nu_1\approx 2$ and $W=7, \nu_1\approx 10$
when the system size $L=6$, but the study of the size dependence of the
conductance $g(L)$ gives $\beta(g)=d log(g) /d log(L) \approx 0$ for W=4.5
(mobility edge in the thermodynamic limit).

 To complete the Coulomb gas model, one needs to know the one--body
potential $V(\lambda)$. In figure 3, $V(\lambda)$ obtained from an
ensemble of numerically calculated
series of $\{\lambda_a\}$ is given. Assuming the mean--field expression
(5), we have directly performed the ensemble average of
$\sum_{a=1}^N \ln |\lambda -\lambda_a|$, that we compare to an analytical
fit whose the functional form is derived in appendix 2:
\be
V(\lambda)=a.\ln^2(1+{\lambda \over \lambda_0}).
\ee
The parameters $a$ and $b=1/\lambda_0$ are adjusted to optimal values
given in figure 3.

\vspace{3ex}
\noindent{\bf 5.  Short range repulsion of the transmission modes and
sensitivity to an applied magnetic field. }
\vspace{1ex}

 The Coulomb gas model implies that successive $\{\nu_a\}$ interact as:
\be
\ln | \cosh(\nu_a) - \cosh(\nu_b)| \approx \ln |\nu_b - \nu_a|
\ee
if they are large and if their spacing is small. This logarithmic
interaction must prevent
small spacings (level repulsion) and the Wigner--surmise
should give an approximate fit for the distribution of their
spacing measured
in average spacing units. We have studied this
distribution, both for the first spacing ($\nu_2-\nu_1$) and for spacings
between $\nu$--levels in the bulk of the spectrum.
 As shown in figure 4, level repulsion persists even in the presence
of transverse localization. For $<\nu_1> \approx 2$
the obtained histogram is very close to the usual prediction of random
matrix theory (Wigner surmise for $\beta=1$). This distribution also
describes the spacings in the bulk of the spectrum (e. g. $\nu_7 -\nu_6$)
where the associated decay lengths $2L /\nu_a$ are quite small.
For larger disorder ( $W=7$, $<\nu_1> \approx 10$), level repulsion is
weaker. The obtained histogram
can then be fitted by an empirical formula proposed by Brody~[24]
(with $q=0.7$)
\be
P_q(s)=As^{q} \exp [-\alpha s^{q+1}]
\ee
in order to interpolate from the correlated random matrix spectra
(Wigner surmise $q=1$) to uncorrelated levels (Poisson distribution
 $q=0$). We can see in figure 2 that the mean spacing
$\nu_2-\nu_1\approx 4$, which can partly explain the weakness of the
repulsion ($\ln |\cosh(\nu_2)-\cosh(\nu_1)|$ is more sensitive to the
large value of $\nu_2$ than to the value of $\nu_2-\nu_1$). However, looking
deeper in the spectrum, we can still note some deviation of the actual
distribution of ($\nu_7-\nu_6$) from the Wigner surmise (figure 5), while
the mean spacing is much smaller. But the persistence of a strong eigenvalue
repulsion between transmission modes with extremely short characteristic
decay lengths ($\xi_7=2L/\nu_7\approx0.5$) is by itself quite noticeable:
the spacing distribution being much closer from a Wigner surmise than from
a Poisson distribution characterizing uncorrelated eigenvalues.

 Since we have approximately obtained spacing distributions in the presence
of time reversal symmetry (orthogonal case) close to the Wigner surmise,
it is interesting to know what is the value of the cross--over field $B^*$
needed for changing this distribution to the Wigner surmise
characteristic of the unitary ensemble. It is also particularly interesting
to see if $B^*$ characterises also the typical field scale where a  change
of $\xi$ occurs, in agreement with our theory relating the value $\xi$
to the importance of the
fluctuation of the $\{\lambda_ a\}$ which are controlled by the ``temperature''
$\beta^{-1}$. We have applied increasing magnetic field  $B$ (measured in
units of $\Phi_o$ per lattice cell).

 We first note that weak fields (about one flux quantum through the sample)
have no clear effects on the different $<\nu_a>$ and on the spacing
fluctuations. Near the mobility edge ($W=4$), we observe as expected
both the Wigner--surmise of the unitary ensemble ($\beta=2$, figure 6)
and a slight change of $\xi$ (increase by a factor $\approx 1.3$) for
$B=0.25$. One can see also in figure 6
that $P(\nu_2 - \nu_1)$ does not depend very much on $B$ for larger disorder.
We think that this is partly due to the gauge invariance of a lattice model
which makes impossible to remove the coherent backscattering effects within
too small localization domains.  Two conclusions emerge from the magnetic field
dependence of the spacing fluctuations. First, the cross--over field $B^*$ is
much
higher in insulators than in metals. Contrary to a metal where the typical
electronic
trajectories are random walks exploring the whole coherent sample, the typical
trajectories in an insulator are enclosing much smaller areas characterized by
much larger dephazing fields. Our results agree with the criterion
$B^* \approx \Phi_0 /\xi^2$. Second, the necessary field in order to
significantly
change $\xi$ is precisely the same which changes the spacing distribution,
in agreement with the existence of a relation between the change of $\xi$
and the change of $\beta$.

 The study of the spacing distribution is not the more accurate test
of the validity of the random matrix ansatz for $3d$--insulators, probing more
the short range eigenvalue repulsion than the long range spectral rigidity.
The usual statistical measure of the long range rigidity is the
$\Delta_3$--statistics introduced by Dyson and Mehta~[25]. Once the spectra
have been numerically unfolded~[26] to series of constant averaged density,
this
statistics measure how the rescaled series deviate from a regular sequence
as a
function of the average number of considered levels. We show in figure 7 how
the
$\Delta_3$--statistics calculated from the $\{\lambda_a\}$ of a $3d$ Anderson
model deviates from the random matrix behaviour when $\xi \approx 1$.
Similar behaviours characterize
the network of random quantum wires previously considered.

 The deviation of $P(S)$ and of
the $\Delta_3$--statistics to the usual random matrix prediction
can be partly due to the highly non--uniform density of the relevant
variables $\{\lambda_a\}$ which could strongly weaken the effect
of the logarithmic repulsion for the rescaled variables. In the quasi--$1d$
limit for instance, we have shown that the ansatz (8) yields quasi--independent
level fluctuations, giving a $P(S)$ or a $\Delta_3$--statistics closer to
Poisson than to Wigner. In references 27, a crossover in $P(S)$ and
in $\Delta_3$ from Wigner to Poisson is obtained within the random matrix
framework, assuming infinite range logarithmic repulsion and a
one--body potential qualitatively similar to Eq. 21. However, since we have
shown for disordered conductors~[20]
that the $3d$--value of the amplitude of the conductance fluctuations is
contradictory with infinite range logarithmic repulsion for
the $\{\lambda_a\}$,
we speculate that the long range part of this repulsion must be even more
severely damped by transverse localization than by transverse
diffusion. But, from the study of those two microscopic models,
we see that a short range repulsion persists in the presence
of transverse localization. The shorter is $\xi$, the shorter is this range.

We note that a dimensionality dependent damping of the
logarithmic repulsion has been shown~[28] to be consistent with the
known density--density correlations of {\it Hamiltonian spectra} in disordered
conductors. Similarly, the long range part of the two body interaction
in (8) may need to be modified in order to still write the distribution
$P(\{\lambda_a\})$ as the Gibbs distribution of a fictitious hamiltonian.

\vspace{3ex}
\noindent{\bf 6.   Disorder dependence of the typical conductance and of its
fluctuations}
\vspace{1ex}

 In this section, we compare the actual behaviour of $<\ln(g)>$ and of
its fluctuations to an extension of the results of section 3 and to
the predictions implied by a direct path approximation~[6-10]. We first note
that contrary to the quasi--$1d$ limit where $-\ln(g) \approx \nu_1$,
$-\ln(g)$ is slightly larger than $\nu_1$ in three dimensions ( $\nu_2,
\nu_3, \ldots$ weakly contribute to $\sum_{a=1}^N (\cosh (\nu_a) -2)^{-1}$).
Extracting the localization length from the size dependence of $<\ln(g)>$
and locating the critical value $W_c$ from the criterion $d \ln(g) /d \ln
(L)=0$
we obtain (figure 8) a simple linear dependence:
\be
\xi^{-1}(W)\propto (W-W_c)
\ee
of the inverse localization length for the $3d$--network of quantum wires.
In the vicinity of the critical point, this linear law is consistent with
the value reported in the experimental litterature for the critical
exponent, but lower than most of the
other numerical results given by finite size scaling studies of various
microscopic models (for a review, see Ref. 29).
But we mainly point out in this study that this linear dependence persists
when the localization length reduces to the lattice spacing, without showing
a logarithmic behaviour that one could naively guess from the length
dependence $h^L$ of the transfer term in a directed path approximation (Eq. 4).

 Considering the sample to sample fluctuations of $\ln(g)$, we have obtained
histograms in rough agreement with a normal distribution of the
log--conductance.
The variance of $\ln(g)$ varies as a function of the disorder and of system
size such that relation (20) continue to be observed (figure 9) in three
dimensions. A similar result have been noticed in the $2d$--localized
regime~[21]
and is consistent with one--parameter scaling, and inconsistent with directed
path theories for the investigated range of parameters.
This one--parameter law is observed in different microscopic models (Anderson
model and network of random quantum wires).
The comparison between the ``random matrix prediction'' and the directed path
result can be sligthly biased since we consider the side to side transmission
through a disorder cube, and not how electrons are transmitted from a corner
to the opposite corner of a cube. We have noted also that the equality between
the mean and the variance of $\nu_1$ is even more accurate than for $\ln(g)$.

\vspace{3ex}
\noindent{\bf 7. Magnetoconductance Fluctuations and Sample to Sample
Fluctuations.}
\vspace{1ex}

 In disordered conductors, a magnetic field $B. L^2 \approx \Phi_0$
removes time-reversal symmetry (cross--over from the orthogonal
ensemble to the unitary ensemble), suppressing weak--localization
corrections and halving the variance of the universal conductance
fluctuations. A (typically weak) magnetic field change of this order
induces also magnetoconductance fluctuations of magnitude equal
(to first order in $1/g$) to the sample to sample fluctuations
under applied magnetic field~[4]. We have already noticed~[15] in two
dimensions
that this ergodicity remains essentially correct in the localized regime
for (large) field change $\Delta B \approx B^*=\Phi_0 / \xi^2$,
at least when $- log(g) \approx 2- 7$. This ``ergodic''
behavior was proposed for explaining the large reproducible
magnetoconductance fluctuations observed in the Mott hopping regime
even for macroscopic size insulators ($L \approx 1 mm$). It is then
interesting to compare in three dimensions the typical
magnitude of the magnetoconductance fluctuations
on a given sample to the sample to sample fluctuations at given field.
 We do not mean here a rigorous statement about possible
exact ergodicity in insulators but a mere comparison of the typical
orders of magnitude. Due to the gauge invariance of our system under
change of applied flux through lattice cell by $\Phi_o$ and due to the
large value of $B^*$ in insulators, a large statistics of those
magnetofluctuations cannot be obtained.

 In Figure 10, we show how  the first $\{\nu_a\}$ ($a=1, 2, \ldots, 5$)
fluctuate for a given sample with $W=5.5$ as a function of $B$.
The fluctuations of $\nu_1$ essentially agree with an ergodic
behaviour, as illustrated in figure 11. In addition to the large
ergodic fluctuations induced by $\Delta B \approx B^*$, one can notice
smaller non ergodic fluctuations which characterizes also the other
$\nu_a$ deeper in the spectrum (figure 10) or $\nu_1$ itself for larger
disorder (figure 12).  We note that larger disorder increases
the scale $B^*$ of the ``ergodic'' fluctuations and prevents us to
observe them when $\nu_1 \gg 1$, leaving only smaller fluctuations
of characteristic field scale which does not seem to depend on $W$.

We point out that $B^*$ in a conductor is temperature dependent
(the relevant area being the coherent domain), while $B^*$
in an insulator corresponds to applying a flux $\approx \Phi_0$
through a temperature independent localization domain typically
smaller than the coherence length $\approx L_{Mott}$ in an activated
regime.

\vspace{3ex}
\noindent{\bf 8.  Effect of a variation of the Fermi energy}
\vspace{1ex}

 Conductance fluctuations can be observed in insulating wires where
the carrier concentration can be controlled by a gate potential, as
shown in Ref. 18. It is interesting to compare the ``gate''--induced
fluctuations observed at very low temperature with the Fermi energy
dependence of $g$ occuring in our microscopic model. Moreover, this
energy dependence  provides interesting information about characteristic
times of the transmission processes. The simpler
case is an isolated resonance describable by a Breit--Wigner formula~[30]:
\be
g(E)\propto T(E) \approx {(\Gamma_e /2)^2 \over {(E-E_r)^2 + (\Gamma_e /2)^2}}
\ee
where $E_r$ is the energy of the resonant state and the inverse of
the resonance width $\Gamma_e$ gives the (long) time spent by the electron
trapped in this resonant localized state.
In more complex cases where resonances and
direct processes are mixed, a smooth energy dependence of the transmission
fluctuations still corresponds to predominant fast direct processes,
while short energy scale fluctuations are related to slow resonant
transmission, a priori neglected in a directed path approximation.
 In figures 13.a and 13.b, one can see how the ``gate''--induced
fluctuations are different when $g \approx 1$
(mobility edge, roughly symmetric ``normal'' fluctuations) and when
the system is deeper in the localized regime (a few resonant peaks
superimposed on a smooth exponentially small background). Both in the
experiments reported in Ref. 18 and here, these peaks can be fitted by
the Breit--Wigner formula (25). Of course in the experiments, inelastic
processes must in general be taken into account~[17].
 In figure 14, we can see how the applied magnetic
field modifies this resonant behaviour. When there is a resonant peak,
a variation of $B$ detunes the resonance from its $B=0$ value, inducing
large conductance fluctuations around that peak. On the contrary, far from
the resonances, in the ``valleys'',  the $B$--dependence is much weaker.
This leads us to distinguish two effects of a variation of $B$. A (small)
orbital effect which modifies the quantum interferences of the shortest
directed paths which controlled a long range hop without intermediary
resonances and a large effect resulting from the shifts of the resonant
peaks. When the sample to sample fluctuations are studied, these two efects
are mixed, though the probability to have a tuned resonance in a sample
taken at random is very small.

A more systematic study of the Wigner time delay matrix will be reported
in a future publication for separating the effects of the long resonances
from the directed path contribution.

\vspace{3ex}
\noindent{\bf 9.  Comparison with level statistics of the Hamiltonian matrix.}
\vspace{1ex}

 In this study, we have investigated the validity of a (too) simple
maximum entropy description of the transfer matrix in the limit of
extreme (transverse) localization. In this description, the
logarithmic interaction between eigenvalues and the value of $\beta$
are related to a good randomization of the eigenvectors. Therefore,
such a description is not appropriate for matrices whose
eigenvector distribution undergoes important change as a function
of the system parameters (e.g. eigenvector localization).

For the hamiltonian matrix, the understanding which emerges from recent
studies~[28, 31] can be summarized as follows in three dimensions. The
logarithmic
repulsion remains valid for
a $N \propto L$ nearest neighbour levels in the metallic phase, for a cube of
size $L^3$. In the localized phase, some level repulsion is still
observed for small $L$, but is expected to disappear in the thermodynamic
limit where the levels would be totally uncorrelated as those of hamiltonian
(1)
in the absence of hopping terms. One can argue that there is a poor
overlap of the wave fuctions due to exponential
localization, such that levels mainly depend on spatially disconnected
part of the random substrate. Therefore levels are less and less correlated
for large $L$ in the absence of long-range correlation of the random
variable $\epsilon_i$ in (1). For finite size, one has intermediary
cases where the spacing distribution $P(S)$ differs from the Wigner surmise
more or less as in figure 4, and the deviations scale with $L$ according to
usual finite size scaling. The mobility edge is characterized~[32] by a novel
scale invariant $P(S)$.

It is then clear that, in the case of the hamiltonian
matrix, ``classical'' random matrix theory with eigenvector distribution
invariant under orthogonal, unitary or symplectic transformations and
without eigenvector--eigenvalue correlations have to be extended to more
complicated theories. Banded matrices with independent entries could be
more realistic for describing hamiltonians in the quasi--$1d$ limit,
but very few analytical results are available for
their eigenvalue distributions~[33].

 The random matrix $X$ differs from the hamiltonian matrix. Localization
appears in the presence of exponentially large eigenvalues
$\{\lambda_a\}$. Our simple maximum entropy approach contains,
via the average density $\sigma(\lambda)$, some information on the
localization of the wave--functions, contrary to a similar approach
for the hamiltonian matrix. One can then expect a better approximation
of spectral correlation functions from this type of approach for $X$
than for $H$. Nevertheless, a numerical study of the $X$--eigenvectors
and of the scale dependence of $P(S)$ in the localized regime will be
very useful.  The persistence of level repulsion for $X$ is associated
with the spatial overlap of electronic paths associated with neighboring
channels. This overlap can be partly related to the importance of
resonant tunnelling through a few localized states. Electronic modes
described by the first few $\nu_a$ would be essentially
modes where the electrons would be trapped
during a long time in overlapping sets of
certain favorable resonant states before being transmitted, as indicated by
the large energy fluctuations shown in section 6. Single scatterer motions
in those sensitive regions would induce
correlated fluctuations of this set of $\nu_a$, as observed in Ref. 12.
On the contrary, in a situation where shortest directed paths are predominant,
one can imagine that the modes of transmission from on side to the opposite
side
of a large cube are associated to different non intersecting paths and
are essentially uncorrelated.

\vspace{3ex}
\noindent{\bf 10.  Comparison with experiments and activated transport.}
\vspace{1ex}

 A comparison between random matrix description, directed path theory,
numerical simulations and low temperature experiments has to be done with
great care.  We consider in this work quantum transmission without activated
process and we do not a priori neglect the important phenomenon of resonant
tunneling which could be more or less truncated at finite temperature
by activated hopping. We refer for this to the join paper by
Bouchaud and Ladieu~[17], where the cross-over between quantum coherent
transmission and activated transport is considered. Let us mention the
physical argument implicitely behind the directed path approximation:
if transport is not entirely coherent, but partly activated, one is precisely
interested by long quantum hop between nearly resonant states, {\it without
additional resonances between them}. In other words, the lifetime of an elastic
resonance could be too long compare to the phase--breaking time at finite
temperature.
Then the propagator evaluated in directed path approximations is implicitely
related to activated transport and might not necessarily describe zero
temperature elastic quantum transmission. In one dimension for instance,
Azbel has pointed out the role
of resonances~[34] and shown clearly that the shortest chains have not
necessarily
the best transmission ( a longer chain with a tuned resonance can be more
transparent for the electrons than a shorter one).

 When the localization length is very short (of order of the Bohr radius of
donor or acceptor impurities), we are dealing with an ``usual'' bounded
state and long range quantum transmission over $L_{Mott}$ is restricted to
interferences between forward scattered paths in a ``cigare shape'' domain
of longitudinal length $L_{Mott}$ and of tranverse length
$(L_{Mott}.\xi)^{\zeta}$.
Those directed hops are characterized by a relatively short time.
When the localization length is longer, we are dealing with a more complicated
Anderson insulator, and one can question if the long time elastic transmission
that we study in this work can occur between successive inelastic collisions.

However, despite the difficulty to take into account activated process and
complicated quantum resonances, an element of answer is given by some
striking similarities between the energy and magnetic field dependence of
the conductance of our zero temperature microscopic model and
the behaviours observed by Ladieu, Mailly, and Sanquer on
a Si:GaAs wire at very low temperature, as illustrated in figure 15.

\vspace{3ex}
\noindent{\bf Appendix 1: Microscopic $3d$--model of disordered quantum
wire networks}
\vspace{1ex}

The studied microscopic model can be considered as a system of narrow wave
guides which intersect each other. We take the limit in which the transverse
size
 of each wave guide tends to zero so that the motion between intersections is
virtually one dimensional. In this limit, there is no underlying simple
Hamiltonian. The mathematical foundation of the model has been studied
by Exner and Seba~[35].
The basic symmetry and conservation laws of quantum mechanics
determine the motion completely once the matching conditions at each
intersection
are specified. Similar concepts have been used by Avron and coworkers in their
study of adiabatic networks~[36]. For simplicity of presentation the number of
space
dimensions is fixed at $d=2$ for the illustrative part (figures etc). Our
numerical
calculations are however carried out for a three dimensional network $d=3$.
Consider then a  square lattice of one dimensional wires with lattice spacing
$a$.
We will denote by $L$ the longitudal dimension of the system and by $N$ its
transverse dimension
(our numerical calculations are performed on cubes with $L=N$).
On this lattice of wires an electron can move subject to the laws of quantum
mechanics. In
the most general case there is a wave number $k_j$ on each bond $j$ of the
lattice.
We have $k_j=\sqrt{E_f-v_j}$ where $v_j$ are random numbers drawn from a
rectangular
probability of width $W$. If $W>E_f$, some of the $k_j$ are purely imaginary
and there
is a disordered combination of plane--wave propagation and tunneling.
       In Fig. (16) we depict such a finite lattice. A site can therefore be
labeled
by a pair of numbers (${\ell},n$), ${\ell}=1,2,\ldots,L; n=1,2,\ldots,N$.
An electron can travel freely on each bond (wire) subject to the Schrodinger
equation.
We will employ the index j for a general bond (horizontal or perpendicular) and
n for horizontal bonds
only. Thus, if the wave number on a certain bond j (j=1,2,..LN) is $k_j$, the
electron's wave function on this bond is a linear combination of plane waves
\be
 	\Psi_j=a_je^{ik_jx}+b_je^{-ik_jx},
\ee
where for horizontal (vertical) bonds, the coordinate x is measured from the
left (down) site. Imagine now that the $N$ sites on the left of the lattice
(these
are the sites $(1,n)$) and the $N$ sites on the right of the lattice (these are
the
sites ($L,n$)) are connected to electron reservoirs by free wires, namely, the
electron's wave number on these wires is the Fermi wave number $k_F$. Then
we have a quantum mechanical  system whose conductance $g$ (at zero
temperature) can
be defined in the following way. Let an incoming wave $e^{ikx}$ of unit
amplitude reach
a site ($1,n$) on the left column of the lattice. Then the wave function at the
$m^{th}$
exit wire on the right (namely to the right of site ($L,m$)) will be
$t_{mn}e^{ikx}$ where
$t_{mn}$ are the elements of the complex transmission matrix $t$. Likewise, the
wave
function at the $m^{th}$ exit wire on the left (namely to the left of site
($1,m$)) will
be $r_{mn}e^{-ikx}$ where $r_{mn}$ are the elements of the complex reflection
matrix $r$.
Unitarity (current conservation) implies the equality
\be
	tt^{\dag}+rr^{\dag}=t^{\dag}t+r^{\dag}r=I_N,
\ee
where $I_N$ is the unit NxN matrix.
	The linear response theory now relates $g$ to the transmission coefficients in
a
simple way. If we consider this system as a collection of random resistances we
notice
that, unlike the  classical situation in which conductances combine rationally
to one another (either in parallel or in series) the situation here is much
more
complicated, since at each  intersection the wave function must be matched
according to the rules of continuity and current conservation. Let $\Psi_1$,
$\Psi_2$, $\Psi_3$ and
$\Psi_4$ be the wave functions of the electron in the four links connected by a
given
site. The four respective wave numbers are denoted by $k_1$, $k_2$,  $k_3$ and
$k_4$
respectively. Then at this site, the following continuity equation should be
satisfied,
\be
	\Psi_1= \Psi_2 =\Psi_3 =\Psi_4  .
\ee
The relation between derivatives at the intersection is somewhat arbitrary as
long as it is compatible with current conservation. For simplicity and
consistence
with previous works~[37-38] we will assume the following relation,
\be
	\Psi_1'+ \Psi_2' =\Psi_3' +\Psi_4'
\ee
where, following Eq. (26),
\be
	\Psi_j'= k_j(a_je^{ik_jx} -b_je^{-ik_jx}).
\ee
These relations give four equations relating the coefficients $a_j$,$b_j$ of
Eq. 26
of the plane waves on each of the four links (see Fig. 16).
	The most natural algorithm by which one can evaluate the conductance is the
transfer
matrix method. Here, however, we intend to study the system also in the far
insulating regime.
A straightforward transfer matrix approach is hence numerically dangerous since
the product
of many matrices may blow up. Furthermore, we intend to study a cubic system
with $L$ up to 12.
For a $3d$--cube ($N=L$) the size of the (complex) transfer matrix will then be
$2N^2=288$,
which is quite large.  It is then indispensable to invent a method
which does not rely on a (possibly divergent) product of large matrices and at
the same time will
employ matrices of smaller size.   This method is explained below in a
heuristic manner. Its
rigorous justification goes back to the theory of wave guides. It involves the
algebra of
transmission and reflection matrices whose size is $N^2$. Recently it has been
adapted for
studying conductances in the insulating regime~[38]. The disadvantage here is
that each
step requires an inversion procedure instead of a simple multiplication as is
required by
the transfer matrix algorithm. The advantage is that the transmission and
reflection matrices are bounded through the unitarity relation and hence there
is no danger
of spurious exponential Lyapunov divergence.

Consider then a wave approaching a system of two barriers ``1'' and ``2''. The
transmission and reflection amplitudes for each individual barrier $i (i=1,2)$,
independent of the other one, are assumed to be given by $t_i$ and $r_i$ if
the wave approaches the barrier from the left and by $t'_i$ and $r'_i$ if the
wave approaches the barrier from the right. For a $N$--channel problem, the
transmission and reflection amplitudes are $N*N$--complex matrices.
They are usually arranged in an $S$--matrix pertaining to barrier $i$,
\be
		S_i=\pmatrix{t_i& r'_i\cr r_i& t'_i\cr} , i=1,2.
\ee
The $S$--matrix is unitary: $Si^{\dag}Si=I$, $SiSi^{\dag}=I$ (the unit matrix
in
the appropriate space).
	The transmission and reflection amplitudes pertaining to the combined
system are $t_{12}$, $r_{12}$ $ t'_{12}$ and $r'_{12}$.
 Our task is then to express these four amplitudes in terms of the amplitudes
of the individual channels.  These relations are given explicitly as follows,
\be
		t_{12}=t_2(1-r'_1r_2)^{-1}t_1
\ee
\be
		r_{12}=r_1+t'_1r_2(1-r'_1r_2)^{-1}t_1
\ee
\be
		t'_{12}=t'_1(1-r_2r'_1)^{-1}t'_2
\ee
\be
		r'_{12}=r'_2+t_2r'_1(1-r_2r'_1)^{-1}t'_2 .
\ee
  In terms of the S matrices  we then write
\be
		S_{12}=S_1*S_2,
\ee
where the star product operation defined through Eqs. (32 - 36) has been
applied by
Redheffer in the theory of waveguides~[39]. It is then possible to add the
barriers
one after the other
until the entire system is worked out. The unitarity bound on the reflection
and
transmission matrices assures the existence of the inverse $(1-r'_1r_2)^{-1}$
and
guarantees the  convergence of the  procedure for any number of steps. It is
worth mentioning here that if the barriers are infinitesimally close to each
other,
the star product is  transformed into a non linear differential relation which
eventually
leads to a matrix Riccati equation for the reflection amplitudes.

We have thus explained how the transmissions of individual units are
combined to give the transmission of
the whole system. Now we have to explain
what are the pertinent units and how the transmission and reflection
through each unit is computed.
Transmission and reflection occur at each column of sites
followed by the accumulation of  phases  as the wave
propagates between two such columns. Thus, each unit consists of a
column of sites together with the links located
on its right side (see Fig. 16). In
order to evaluate the transmission and reflection at a given column of
sites we number the sites from 1 to $N$, and consider a site $m
(m=1,2,...N)$. The wave number to its left is $k_m$ while the wave
number to its right is $p_m$. The wave numbers in the transverse
direction are $q_m$ (above the site) and $q_{m-1}$ (below the site).
We adopt here periodic transverse boundary conditions
so that $q_0=q_N$. On a transverse link
between sites $m$ and $m+1$ the wave function is given by
\be
		\Psi(y)=c_me^{iq_my}+d_me^{-iq_my}  ,
\ee
 where the coordinate $y$ is measured from link $m$.
	An incoming wave of unit amplitude reaching site $n$
from the left will provoke, in any site $m$, leftward reflected
waves with reflection amplitude $r_{mn}$, as well as rightward
transmitted waves  with transmission amplitude $t_{mn}$. The
matching equations [Eqs. (28 - 29)] now imply the following relations
(the dependence of the transverse coefficients $c$ and $d$ on the
incoming index $n$ will  not be indicated.
\be
	\delta_{mn}+r_{mn}=t_{mn}=c_m+d_m=c_{m-1}
e^{iq_{m-1}a}+d_{m-1}e^{-iq_{m-1}a},
\ee
\be
	k_m( \delta_{mn}-r_{mn})+q_{m-1}(c_{m-1}
e^{iq_{m-1}a}-d_{m-1}e^{-iq_{m-1}a})=p_mt_{mn}+q_m(c_m-d_m),
\ee
where $\delta_{mn}$ is the Kronecker delta function.
The equality $\delta_{mn}+r_{mn}=t_{mn}$ results simply
since reflection and transmission
occur at the same point in space. From the first set of
equations (39) one can now eliminate the coefficients of the
transverse motions pair by pair using the relations
\be
		c_m+d_m=t_{mn}
\ee
\be   c_me^{iq_ma}+d_me^{-iq_ma}=t_{m+1,n}.
\ee
We now substitute the solution of equations (40 - 41) into equation (39)
 and replace $r_{mn}$ by $t_{mn} - \delta_{mn}$.
 As a result we get a set of equations for the transmission matrix
$t={t_{mn}}$ of the pertinent site of columns which reads
\be
		At=2K,
\ee
where the $N*N$--matrices $A$ and $K$ are given by
\be
		A(m,m)=k_m+p_m+i[q_m\cot(q_ma)+q_{m-1}\cot(q_{m-1}a)]
\ee
\be
		A(m,m+1)=-iq_m\csc(q_ma)
\ee
\be
		A(m,m-1)=-iq_{m-1}\csc(q_{m-1}a)
\ee
\be
		K=diag(k_1,k_2,...k_N) ,
\ee
with $m=1,2,\ldots,N$ and $N+1$ or $0$ are defined modulo $N$.
The solution of Eq. (42) yields the transmission matrix from left to right
 and then the corresponding left to left reflection matrix $r=t-I_N$ ($I_N$
is the unit matrix) is obtained. To get the matrices pertaining to waves
approaching from the right, one has to interchange the roles of $k_m$ and
$p_m$.
Since the matrix $A$ is not affected by this swap,
it does not require any additional matrix inversion procedure.
	It is worthwhile noticing here that in the presence of a constant
magnetic field $B$ perpendicular to the direction of
propagation the only change in the above formalism concerns the non diagonal
elements of the matrix $A$. If the columns of
sites are numbered as $1,2,\ldots,L$, and the magnetic flux per plaquette
is $\phi=Ba^2$ in units of $\Phi_0$) then at column ${\ell}$
(${\ell}=1,2,\ldots,L$)
one has
\be
		A(m,m+1)\to A(m,m+1)e^{2\pi i{\ell}\phi},
\ee
\be
		A(m,m-1)\to A(m,m-1)e^{-2\pi i{\ell}\phi}.
\ee
These phases are identical to those multiplying the hopping integrals
(the Peierls substitution) used e.g in  tight binding models.
	What is left in order to complete the calculations of the individual
unit shown in Fig. (16) is to take account of the
phase between two units. This is easily accomplished using the matrix
P=diag($p_1,p_2,...p_N$) and the substitutions
\be
		t\to e^{iP}t, r\to r, t'\to t'e^{iP},  r'\to e^{iP}r'e^{iP}.
\ee
	We have thus explained how to evaluate the transmission and
reflection amplitudes for each individual unit in the random
link model. Together with the composition law explained in the first
part of this section (Eqs. (33 - 36)) we have at hand an algorithm
for a numerical study of the full system. We repeat here that
our approach will mostly be used in  three dimensions, for which
 the relevant equations are slightly more complicated.

The radial parameters $\{\lambda_a\}$ are then obtained from the eigenvalues
$T_a$ of ${\bf t.t}^{\dagger}$ using $T_a=1/(1+\lambda_a)$.

\vspace{3ex}
\noindent{\bf Appendix 2. Functional form of the analytical fit for
$V(\lambda)$}
\vspace{1ex}

 Although the actual density $\sigma(\nu)$ differs from a simple
uniform density between two nonzero values, it is interesting to
to investigate what $V(\lambda)$ is yielded from this crude
approximation (a uniform density
for the $\nu$--variable between two strictly positive values $2L/A$
and $2L/B$). One gets:
\be
U(\nu)=C. \int_{2L/A}^{2L/B} ln |\cosh (\nu)-\cosh (\nu ')| d\nu '
\ee
where $C=(N / 2L)(1/B-1 /A)^{-1}$.

The derivative is given by:
\be
{\partial \over \partial \nu}U(\nu)=C.\ln |({\exp(2L/A)-\exp( \nu) \over
\exp(2L/ A) - \exp -(\nu)}). ({\exp(2L/B)-\exp -(\nu) \over exp (2L/B)
- \exp (\nu)})|
\ee
which presents spurious divergences at the band edges $2L/A$ and $2L/B$.
Ignoring those divergences and after integration, one can see that
the corresponding $U(\nu)$ is characterized in presence of transverse
localization ($A \neq 0$) by a flat part for
$0 \leq \nu \leq \nu_0 \approx 2L/A$
and by a quadratic increases $\propto (\nu -\nu_0)^2$ for $\nu_0 \leq \nu$.
Coming back to the original variable $\lambda$, we then expect a
confining potential increasing as $(\ln(\lambda)-\ln(\lambda_0))^2$ for
$\lambda \gg \lambda_0 \gg 1$ and presenting a flat part for $\lambda \leq
\lambda_0$. This suggests that we try to fit the $V(\lambda)$ calculated from
a microscopic $3d$--model assuming relation (5) with the following
analytical expression
\be
V(\lambda)=a.\ln^2(1+{\lambda \over \lambda_0}).
\ee
The agreement is good for small enough values of $\lambda$, as shown in
figure 3. It is worth to notice that the validity of this fit seems very
general: it has been
checked in $2d$ and $3d$ disordered metals modelled either by random
quantum wire networks or by tight--binding Anderson hamiltonians~[20].

\vspace{3ex}
\noindent{\bf Acknowledgements}
\vspace{1ex}

J.L.P. has benefited from useful discussions with the other members of
the 1991--program on ``mesoscopic system'' at the Institute for Theoretical
Physics at Santa Barbara and with the members of the E.E.C. Science program
Contract No. SCC-CT90-0020 ``Localization and Transport Fluctuations in
Microstructures'', notably B. Kramer and P. Markos. We acknowledge also
extensive discussions with J.P. Bouchaud, F. Ladieu and M. Sanquer.
Part of this work at University of Florida was supported by N.S.F.
grant no. DMR--8813402.

\newpage

\vspace{3ex}
\noindent{\bf Figures Captions}
\vspace{1ex}

\vspace{1ex}
Figure 1: Average density $\sigma(\nu)$ calculated for cubes of size $L=6$
for disorder parameters $W=4$ (circles) and $W=7$ (squares).

\vspace{1ex}
Figure 2:  Ensemble averaged $<\nu_n>$, $n=1, \ldots , L^2$ for the
same cubes as in Fig. 1.

\vspace{1ex}
Figure 3: Potential $V(\lambda)$ calculated for cubes of size $L=8$
for different values of $W$ (squares) and the analytical fit given
in Eq. 21 for indicated values of a and $b=\lambda_0^{-1}$.

\vspace{1ex}
Figure 4: $P(S)$ for $S=(\nu_2-\nu_1)/<\nu_2-\nu_1>$ corresponding
to the same cubes as in Fig 1-2 for $W=4$ (circles), $W=7$ (squares)
and Brody distribution with $q=0.7$ (thick line).

\vspace{1ex}
Figure 5: $P(S)$ for $S=(\nu_7-\nu_6)/<\nu_7-\nu_6>$ for $L=6, W=7$
compared to the Wigner surmise.

\vspace{1ex}
Figure 6 :  $P(S)$ for $S=(\nu_2-\nu_1)/<\nu_2-\nu_1>$ for cubes of
size $L=8$ $W=4$ (circles $<\nu_1>\approx2.02$) and $W=7$ (Squares
$<\nu_1> \approx 10.97$) compared with the Wigner surmise for
$\beta=1, 2$. Applied magnetic field $B=0.25$.

\vspace{1ex}
Figure 7: $\Delta_3$--statistics as a function of
the mean number $n$ of eigenvalues $\lambda_a$. Unfolded spectra
obtained in a $3d$-Anderson-model for $W=30$ and $L=7$ at the band
centre ($\xi \approx 1$). The continuous curves are given by $n/15$
(uncorrelated spectra) and $[\ln(2\pi*n) +\gamma-5/4-\pi^2/8]/\pi^2$
(random matrix theory).

\vspace{1ex}
Figure 8: The inverse localization length as a function of $W-W_c$ for
$W_c = 4.5$. Values of $L$ beween 6 and 12 are used to evaluate $\xi$.

\vspace{1ex}
Figure 9: average (circles) and variance (diamonds) of $\log(g)$ as
a function of $W$ for $3d$ cubes ($L=8$).

\vspace{1ex}
Figure 10: The first eigenvalues $\nu_a$ of a single sample ($L=10$)
at moderate disorder ($W=5.5$) as a function of the applied magnetic field $B$.

\vspace{1ex}
Figure 11: Magnetic field dependence of $\nu_1$ for five different samples
($L=10, W=5.5$).

\vspace{1ex}
Figure 12: Magnetic field dependence of $\nu_1$ for two samples at
different values $W$ ($L=10$).

\vspace{1ex}
Figure 13.a: Fermi energy dependence of the conductance for a single sample
deep in the localized regime ($g \ll 1, L=8, W=6, B=0$).

\vspace{1ex}
Figure 13.b: Fermi energy dependence of $g$ near the mobility edge
($L=8, W=6, B=0$).

\vspace{1ex}
Figure 14: Fermi energy dependence of the log-conductance of
a single insulating sample ($L=8, W=6$) for increasing magnetic
field $B=0, 0.1, 0.2, \ldots, 0.5$.

\vspace{1ex}
Figure 15: Curve A: $log(g)$ as a function of the gate voltage (upper
coordinate) observed in a mesoscopic GaAs:Si wire (Ref.18) where $g$ has
been multiplied by a factor 40. Curve B: numerical simulation of $log(g)$
as a function of $E_F$ using the $3d$--quantum wire network.

\vspace{1ex}
Figure 16: Scheme of the network of quantum wires used as a microscopic model.

\newpage

\vspace{3ex}
\noindent{\bf References} \\

\noindent
[1] E. Abrahams, P. W. Anderson, D. C. Licciardello and T. V. Ramakrishnan,
1979,
 {\it Phys. Rev. Lett.}, {\bf 42}, 673.

\noindent
[2] F. Wegner, 1979, {\it Z. Phys.} B, {\bf 35}, 207.

\noindent
[3] For reviews see {\it Mesoscopic Phenomena in Solids} (North-Holland,
Amsterdam, 1991),
ed. by B. L. Althshuler, P. A. Lee and R. A. Webb.

\noindent
[4] B. L. Altshuler and D. E. Khmelnitskii, 1986, {\it JETP Lett.}, {\bf 42},
359.

\noindent
[5] S. Feng, P. A. Lee and A. D. Stone, 1987, {\it Phys. Rev. Lett.}, {\bf 56},
1960.

\noindent
[6] For reviews see {\it Hopping Transport in Solids} (North Holland,
Amsterdam, 1991), ed.
by M. Pollack and B. Shklovskii. (more particularly chapter 9).

\noindent
[7] E. Medina, M. Kardar, Y. Shapir and X. R. Wang, 1989, {\it Phys. Rev.
Lett.},
{\bf 62}, 941; J. Cook and B. Derrida, 1990, {\it J. of Stat. Phys.},
{\bf 61}, 961.

\noindent
[8] E. Medina and M. Kardar, 1992, {\it Phys. Rev.}, {\bf B 46}, 9984.

\noindent
[9] E. Medina, M. Kardar, Y. Shapir and X. R. Wang, 1990, {\it Phys. Rev.
Lett.},
{\bf 64}, 1816; H. L. Zhao, B. Spivak, M. Gelfand and S. Feng, 1991,
{\it Phys. Rev.}, {\bf B44}, 10760.

\noindent
[10] E. Medina and M. Kardar, 1991, {\it Phys. Rev. Lett.}, {\bf 66}, 3187;
Y. Meir, N. S. Wingreen, O. Entin--Wohlman and B. L. Altshuler, 1991,
{\it Phys. Rev. Lett}, {\bf 66}, 1517.

\noindent
[11] V. L. Nguyen, B. Spivak and B. Shklovskii, 1986, {\it JETP Lett.}, {\bf
43}, 44.

\noindent
[12] S. Feng, J.--L. Pichard and F. Zeng, 1993, preprint.

\noindent
[13] J.--L. Pichard, 1991, in {\it Quantum Coherence in Mesoscopic Systems},
(NATO ASI series, Plenum, New York), ed. by B. Kramer;  A. D. Stone,
P. A. Mello, K. Muttalib and J.--L. Pichard in Ref. 3.

\noindent
[14] J.--L. Pichard, M. Sanquer, K. Slevin and P. Debray, 1990, {\it Phys. Rev.
Lett.},
{\bf 65}, 1812.

\noindent
[15] S. Feng and J.--L. Pichard, 1991, {\it Phys. Rev. Lett}, {\bf 67}, 753.

\noindent
[16] P. Hernandez and M. Sanquer, 1992, {\it Phys. Rev. Lett.}, {\bf 68}, 1402.

\noindent
[17] F. Ladieu and J. P. Bouchaud, companion paper submitted to. J. de
Physique.

\noindent
[18] F. Ladieu, D. Mailly and M. Sanquer,  companion paper submitted to. J. de
Physique.

\noindent
[19] K. Slevin, J.--L. Pichard and P. A. Mello, 1991, {\it Europhys. Lett},
{\bf 16}, 649.

\noindent
[20] K. Slevin, J.--L. Pichard and K. Muttalib, 1993, to appear in {\it J. de
Physique I}.

\noindent
[21] J.--L. Pichard, N. Zanon, Y. Imry and A.D. Stone, 1989, {\it J. de
Physique},
{\bf 51}, 587.

\noindent
[22] P.A Mello, P. Pereyra and N. Kumar, 1988, {\it Ann. of Phys.}, {\bf 181},
290.

\noindent
[23] P. Markos and B. Kramer, 1993, preprint.

\noindent
[24] T. Brody, 1973, {\it Lett. Al Nuovo Cimento}, {\bf 7}, 482.

\noindent
[25] F. Dyson and M. L. Mehta, 1963, {\it J. of Math. Phys.}, {\bf 4}, 701.

\noindent
[26] O. Bohigas, 1991, in {\it Chaos and Quantum Physics} (North Holland,
Amsterdam)
ed. by M.--J. Giannoni, A. Voros and J. Zinn--Justin.

\noindent
[27] Muttalib, Chen, Ismail and Nicopoulos, preprint; Blecken and Muttalib,
unpublished.

\noindent
[28] R. Jalabert, J.--L. Pichard and C.W.J. Beenakker, 1993, preprint.

\noindent
[29] B. Kramer, K. Broderix, A. MacKinnon and M. Schreiber, 1990, {\it
Physics},
{\bf A 167}, 163.

\noindent
[30] A. D. Stone and P. A. Lee, 1985, {\it Phys. Rev. Lett.}, {\bf 54}, 1196.

\noindent
[31] B. L. Altshuler and B. I. Shklovskii, 1986, {\it Sov. Phys. JETP}, {\bf
64}, 127.

\noindent
[32] B. I. Shklovskii, B. Shapiro, B. R. Sears, P. Lambrianides and H. B.
Shore,
1993, to appear in {\it Phys. Rev.}, {\bf B}.

\noindent
[33] G. Casati, F. Izrailev and L. Molinari, 1990, {\it Phys. Rev. Lett.}, {\bf
64},
1851;  M. Feingold, A. Gioletta, F. Izrailev and L Molinari, 1993, to appear
in {\it Phys. Rev. Lett.}

\noindent
[34] M. Azbel, 1983, {\it Solid State Commun.}, {\bf 45}, 527.

\noindent
[35] 0 P. Exner and P. Seba, 1989, {\it Reports on Mathematical Physics},
{\bf 27}, 7.

\noindent
[36] J. E. Avron and L. Sadun, 1989, {\it Phys. Rev. Lett.}, {\bf 62}, 3082.

\noindent
[37] Y. Avishai and Y. B. Band, 1987, {\it Phys. Rev. Lett.}, {\bf 58}, 2251

\noindent
[38] Y. Avishai and J. M. Luck, 1992, {\it Phys. Rev.}, {\bf B 45}, 1074.

\noindent
[39] M. Redheffer, 1962, {\it J. of Math. Phys.}, {\bf 41}, 1.

\end{document}